\newcommand{\chandra}{{\it Chandra}}
\newcommand{\xmm}{{\it XMM}}
\newcommand{\suzaku}{{\it Suzaku}}
\newcommand{\asca}{{\it ASCA}}
\newcommand{\RXJ}{RX~J1713.7$-$3946}

\documentclass{iau}
\usepackage{graphicx}

\title[Thermal X-ray Spectra of SNRs] 
{Thermal X-ray Spectra of Supernova Remnants}

\author[P. Slane] 
{Patrick Slane$^1$}

\affiliation{$^1$Harvard-Smithsonian Center for Astrophysics \\ 60 Garden Street \\
Cambridge, MA 02138 USA \\
email: {\tt slane@cfa.harvard.edu}}

\pubyear{2008}
\volume{296}  
\pagerange{xxx-yyy}
\jname{Supernova Environmental Impacts}
\editors{A. Rey, R. McCray eds.}
\begin{document}

\maketitle

\begin{abstract}
The fast shocks that characterize supernova remnants heat circumstellar
and ejecta material to extremely high temperatures, resulting in
significant X-ray emission. The X-ray spectrum from an SNR carries
a wealth of information about the temperature and ionization state
of the plasma, the density distribution of the postshock material,
and the composition of the ejecta. This, in turn, places strong
constraints on the properties of the progenitor star, the explosive
nucleosynthesis that produced the remnant, the properties of the
environment into which the SNR expands, and the effects of particle
acceleration on its dynamical evolution. Here I present results
from X-ray studies SNRs in various evolutionary states, and highlight
key results inferred from the thermal emission.
\keywords{(ISM:) supernova remnants, X-rays: ISM}
\end{abstract}

\firstsection 
\section{Introduction}
\noindent
The very fast shocks formed in young and middle-aged supernova
remnants (SNRs) act to heat matter to temperatures exceeding many
millions of degrees. As a result, these systems are copious emitters
of thermal X-rays. This emission, characterized by bremsstrahlung
continuum accompanied by line emission from recombination and
de-excitation in the ionized gas, can originate from three distinct
regions of the SNR: (1)behind the forward shock (FS),
where interstellar or circumstellar material is swept-up and heated;
(2)interior to the SNR boundary, where cold ejecta are heated by
the reverse shock (RS); and (3)outside a central pulsar wind nebula
(PWN), if one exists, where the slow-moving central ejecta are
heated by the expanding nebula.  Thermal X-ray spectra in an SNR
thus provide crucial information on the surrounding environment,
which may have been modified by strong stellar winds from the
progenitor, as well as the ejecta that bears the imprint of the
stellar and explosive nucleosynthesis in the progenitor star.

Studies of the thermal X-ray emission from SNRs reveal details on
the temperature, density, composition, and ionization state of the
shocked plasma. These, in turn, can provide specific information
on the density structure for both the ejecta and the surrounding
circumstellar material, the nature of the supernova explosion
(core-collapse vs. Type Ia), the age and explosion energy of the
SNR, the mass of the progenitor star, and the distribution of metals
in the post-explosion supernova.  Here I present a summary of results
from several studies of the thermal spectra of SNRs. This review
is brief; space prohibits an in-depth discussion of many aspects
and results from this rich field. For an excellent review of X-ray
emission from SNRs, including a detailed discussion of the thermal
X-ray emission, the reader is referred to Vink et al. (2012).

\section{X-rays from SNRs}
\noindent
Spectral analysis of thermal X-ray emission can provide
measurements of the electron temperature and the relative abundances
in the shocked plasma. These quantities are important for studies
of the dynamical evolution of the SNR and the composition of the
shocked gas. However, the low density environments in which SNRs
evolve result in crucial effects that affect the interpretation of
the spectra. Of particular importance are the relatively long
timescales for electron-ion temperature equilibration and ionization
equilibrium.

\subsection{Electron-Ion Temperature Equilibration} 
In the simplest picture, and for an ideal gas, all particles that
go through the SNR shock attain a velocity $v = 3 v_s/4$ where $v_s$
is the shock velocity. Since the associated kinetic temperature
scales with mass, the result is that the electron temperature is
initially much lower than that of the ions. Because the ions comprise
the bulk of the mass swept up by the SNR, it is the ion temperature
that characterizes the dynamical evolution. However, the temperature
determined from X-ray measurements is that of the electrons. Thus,
in connecting X-ray measurements to the SNR evolution, modeling of
the electron-ion temperature ratio is required.

On the slowest scales, Coulomb collisions between electrons and
ions will bring the populations into temperature equilibrium. More
rapid plasma processes may result in faster equilibration, and often
the two extreme cases of instantaneous temperature equilibration
and Coulomb equilibration are considered in order to investigate
the boundaries of the problem. SNR expansion velocities can be used
to estimate the ion temperatures for some remnants, and the estimated
temperatures are much higher than the observed electron temperatures,
suggesting slow temperature equilibration.  However, the effects
of cosmic-ray acceleration can result in ion temperatures lower
than are indicated by the shock velocity (see Section 5), making
more direct measurements of both $T_e$ and $T_i$ crucial.

In a small number of cases, measurements of Balmer lines at the
forward shock can provide the proton temperature and the degree of
electron-ion equilibration, and there is a significant trend for
larger $T_e/T_p$ values with lower shock velocities, potentially
consistent with electron heating by lower hybrid waves (Ghavamian
et al. 2007).  Future high-resolution X-ray measurements of thermal
line broadening offer the promise of direct measurements of both
$T_e$ and $T_p$ in SNRs.

\begin{figure}[t]
\begin{center}
 \includegraphics[angle=0,width=1.00\textwidth]{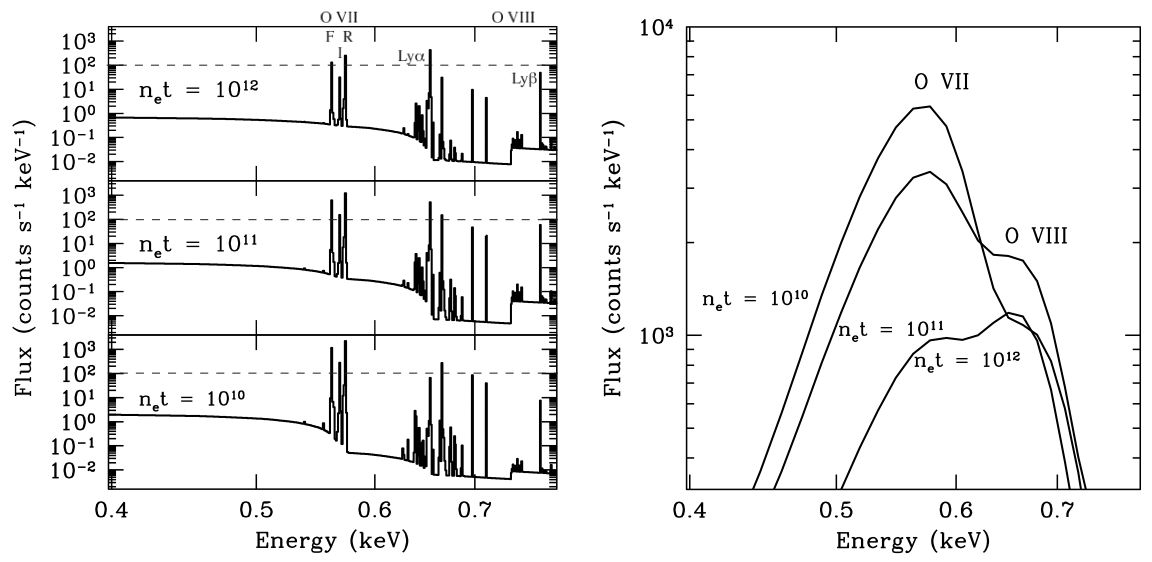}
 \caption{
Left: Evolution of X-ray spectrum with ionization age, emphasizing
lines from O VII and O VIII. The plasma temperature is $10^{6.5}$~K,
and the units for $n_e t$ are ${\rm cm}^{-3}{\rm\ s}^{-1}$. The
dashed line at $10^2 {\rm\ counts\ s}^{-1}{\rm\ keV}^{-1}$ is simply
for comparison purposes.  Right: Spectra at left folded through the
\chandra\ ACIS-S detector response.
}
   \label{fig1}
\end{center}
\end{figure}

\subsection{Ionization Effects} 
Because the density of the X-ray emitting plasma in SNRs is extremely
low, the ionization state of the gas takes considerable time to
reach the equilibrium state that is characteristic of its temperature.
The ionization parameter $\tau = n_e t$ determines how far the
ionization has progressed.  The plasma reaches collisional ionization
equilibrium (CIE) for $\tau \gtrsim 10^{12.5}{\rm\ cm}^{-3}{\rm\
s}$; for smaller values the plasma is in a nonequilibrium ionization
(NEI) state. This is illustrated in Figure 1. The left panel shows
a portion of the soft X-ray spectrum for a plasma at a temperature
of $10^{6.5}$~K for different values of $n_e t$. Of particular note
is the relative strengths of the O~VII triplet and the Ly$\alpha$
and Ly$\beta$ lines from O~VIII.  It is clear from this that
determination of the plasma temperature based on relative line
strengths must thus account for NEI conditions in the plasma.  In
the right panel, we plot the same spectra folded through the \chandra\
ACIS-S detector response. While the limited spectral resolution
provided by CCD detectors obviously obscures details of specific
line ratios, spectral fits to the spectrum can easily identify gross
differences of the ionization state.

The progression of the ionization state in the evolving postshock
gas in SNRs is readily observed. Early \chandra\ observations of
1E~0102.2$-$7219, for example, show that the peak of the O~VII
emission is found at a smaller radius than that of O~VIII, consistent
with the expectation that the ionization state of ejecta most
recently encountered by the inward-propagating RS lags behind that
of the ejecta that have been shocked earlier (Gaetz et al. 2000).
The same effect is evident in shocked circumstellar material in
G292.0$+1.8$, where the ionization state of the emission directly
behind the FS is observed to be considerably lower than that for
regions further downstream (Lee et al. 2010). Such measurements
thus provide constraints on the thermal history of the shocked
material as well as on variations in the density structure.

While conditions of underionization are found in many SNRs, recent
X-ray observations have identified exactly the opposite situation
in several remnants. Kawasaki et al. (2005) used the relative
intensity of H-like and He-like lines of Ar and Ca from \asca\
observations of W49B to determine an ionization temperature that
is higher than $T_e$, indicating an {\it overionized} plasma. Ozawa
et al.  (2009) used \suzaku\ observations to identify a distinct
radiative recombination continuum (RRC) feature that confirms this
overionized state (Figure 2, left). Using \xmm\ observations, Miceli
et al. (2010) find that the overionization in W49B appears to be
related to regions in which rapid expansion and adiabatic cooling
have lowered the plasma temperature to a value below its ionization
state. Similar results have been observed for IC~443 (Yamaguchi et
al. 2009). A semi-quantitative analysis by Moriya (2012) suggests
that the progenitors of such overionized SNRs may be massive RSG
stars for which the wind-driven circumstellar environment is
sufficiently dense to promote rapid early ionization followed by a
drop in temperature as the SNR expands into the lower density regions
of the wind shell. This is illustrated in Figure 2 (right) where
the electron density encountered by a $10,000 {\rm\ km\ s}^{-1}$
shock traveling through a CSM formed by a stellar wind is shown.
The dashed line follows regions for which the plasma will reach
approximate ionization equilibrium. The shaded region corresponds
to mass-loss rates typical of a massive RSG progenitor, and shows
that for early times, the plasma may be left in an overionized
state. Such conditions appear to be connected to the mixed-morphology
class of SNRs, making further study of the dynamical conditions
leading to such overionized states of particular interest.

\begin{figure}[t]
\begin{center}
 \includegraphics[width=1.0\textwidth]{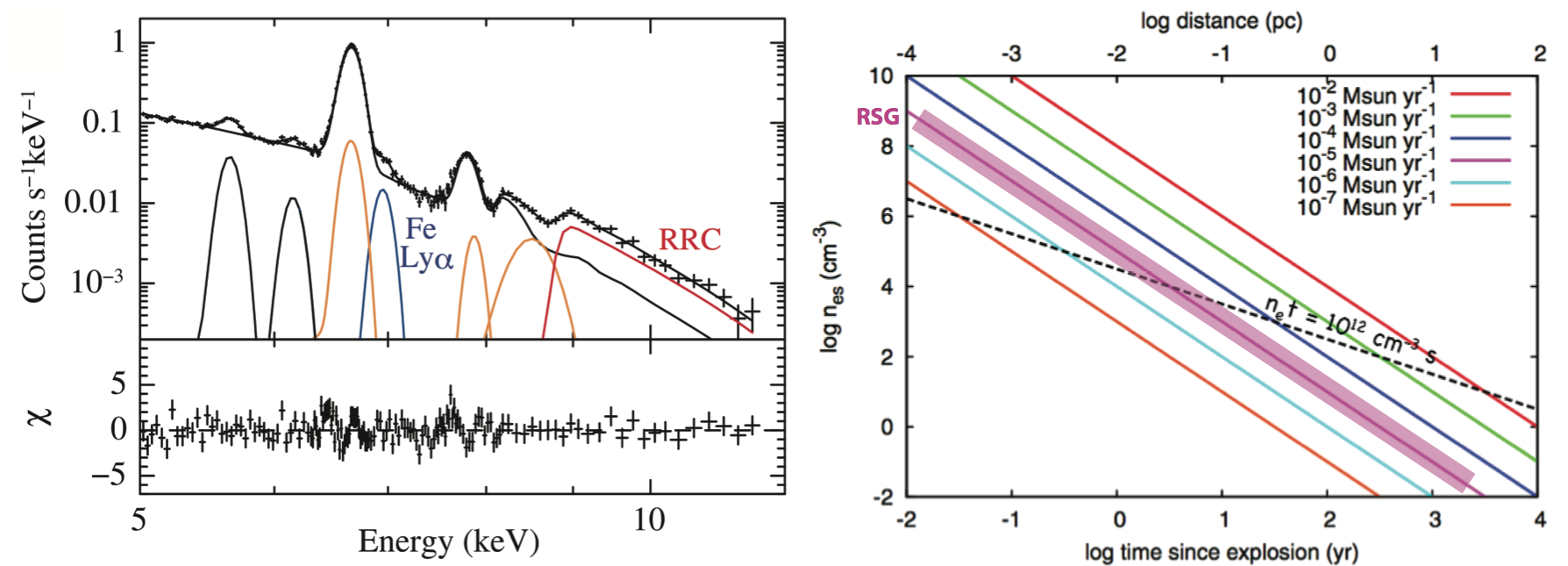}
 \caption{
Left: \suzaku\ spectrum for W49B, showing RRC features and enhanced
H-like emission from Fe, indicative of overionized plasma. (From Ozawa
et al. 2009.)
Right: Electron density distribution for wind model as a function of
time (and distance) traveled by $10,000 {\rm\ km\ s}^{-1}$ SNR shock. 
See text for description. (From Moriya 2012. Reproduced by permission 
of the AAS.)
}
   \label{fig2}
\end{center}
\end{figure}

\section{Studies of SNR Ejecta}
\noindent
The very different stellar evolution histories and explosion processes
for Type Ia and core-collapse (CC) SNe result in distinct signatures
in the shock-heated ejecta of SNRs. Type Ia events, corresponding
to the complete disruption of a C/O white dwarf star, produce more
than $0.5 M_\odot$ of Fe-group elements, accompanied by a significant
contribution of intermediate mass elements. CC SNe, on the other
hand, are dominated by materials synthesized during the stellar
evolution of the massive progenitor -- particularly O -- with
additional products from explosive nucleosynthesis in the innermost
regions surrounding the collapsed core.  As illustrated in Figure
3 (left), where we plot the mass distributions for key nucleosynthesis
products for characteristic Type Ia and CC events (Iwamoto et al.
1999), the former are dominated by Fe while the latter contain much
larger amounts of O. For comparison, the total mass of these elements
contained in $10 M_\odot$ of swept-up material with solar abundances
is also shown. Particularly at young ages when the total amount of
mass swept up by the FS is not exceedingly high, the thermal X-ray
spectra from such remnants provide rich information about supernova
ejecta.

In Figure 3 (right), we compare the spectra from N103B (top), a
Type Ia SNR (Lewis et al. 2003) with that from the CC
SNR G292.0$+1.8$ (bottom). The dominant flux just below 1~keV in
N103B is largely from Fe-L emission, characteristic of the large
amount of Fe created in such events, while the spectrum from
G292.0$+$1.8 shows strong emission features from O and Ne (Park et
al. 2004).  Identification and modeling of such spectral features
has been used to identify the SN type that produced numerous SNRs,
notably Kepler's SNR, which Reynolds et al. (2007) identified as a
Type Ia remnant based on the dominant Si, S, and Fe emission.
Subsequent detection and modeling of the Mn and Cr lines in Kepler
imply a high-metallicity progenitor star (Park et al. 2013).

\begin{figure}[t]
\begin{center}
 \includegraphics[width=1.00\textwidth]{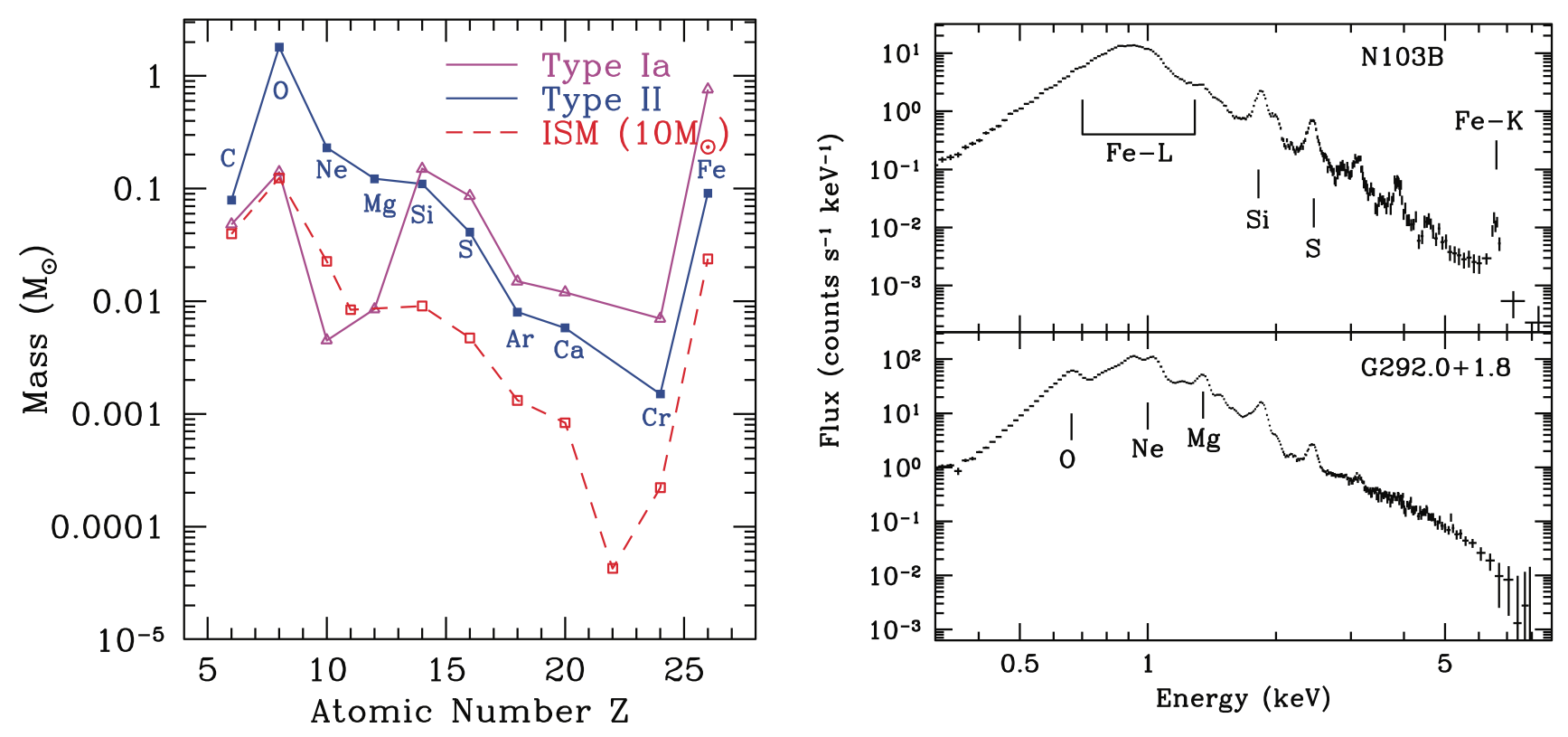}
 \caption{
Left: Mass of ejecta components, by atomic number, in representative
core-collapse and Type Ia SNe. For comparison, the mass contained in
$10M_\odot$ of solar-abundance material is also shown.
Right: Thermal X-ray spectra from a Type Ia remnant (N103B) and
core-collapse remnant (G292.0$+$1.8), illustrating the significant
difference in Fe and O/Ne content.
}
   \label{fig3}
\end{center}
\end{figure}

A particularly important example of SNR ``typing'' through X-ray
spectra is that of SNR 0509$-$67.5. Hughes et al. (1995) used \asca\
spectra to argue for a Type Ia progenitor for this remnant. By
comparing different Type Ia explosion models with spectra from \xmm\
and \chandra\ observations, Badenes et al. (2008) argued that
0509$-$67.5 is the result of an energetic, high-luminosity SN~1991T-like
event, a result subsequently confirmed with light echo spectra from
the original event (Rest et al. 2008).

The early development of SN explosions can imprint signatures on
the SNRs that they form. Studies of the spatial distribution of
ejecta can thus provide evidence of asymmetries and mixing in these
events. X-ray studies of Cas A, for example, show distinct evidence
of Fe ejecta in the outermost regions of the remnant (Hughes et al.
2000a), despite the expectation that Fe is produced in the regions
closest to the remnant core. This large-scale disruption of the
ejecta layers has been studied in detail by Hwang and Laming (2012)
who performed fits to over 6000 X-ray spectra in Cas A and find
distinct examples of regions where Fe is accompanied by other
products of incomplete Si burning, along with others that are nearly
pure Fe, presumably produced in regions of $\alpha$-rich freezeout
during complete Si burning.  They conclude that nearly all of the
Fe in Cas A is found outside the central regions of the remnant,
apparently the result of hydrodynamic instabilities in the explosion.

The large-scale distribution of SNR ejecta has been investigated
for a large number of SNRs by Lopez et al. (2011), who find that
the thermal X-ray emission from remnants of Type Ia events shows a
higher degree of spherical symmetry and mirror symmetry than for
CC remnants. This may indicate that CC SNe
evolve in more asymmetric environments, or perhaps that the events
themselves are asymmetric.

\section{Studies of Shocked Circumstellar Material}
\noindent
Remnants of CC SNe initially evolve in the circumstellar
environment left by their associated progenitors. For progenitors
with significant pre-explosion wind phases, the remnants initially
evolve into density profiles with $\rho \propto r^{-2}$. The composition
of the circumstellar material from progenitors with strong RSG or
WR wind episodes is expected to show the signatures of the CNO cycle,
which leads to an environment with an enhanced N/O ratio. Studies
of thermal X-ray emission from behind the FS can provide
constraints on the shocked CSM, and thus on the late-phase
properties of the progenitor star.

\begin{figure}[t]
\begin{center}
 \includegraphics[width=1.00\textwidth]{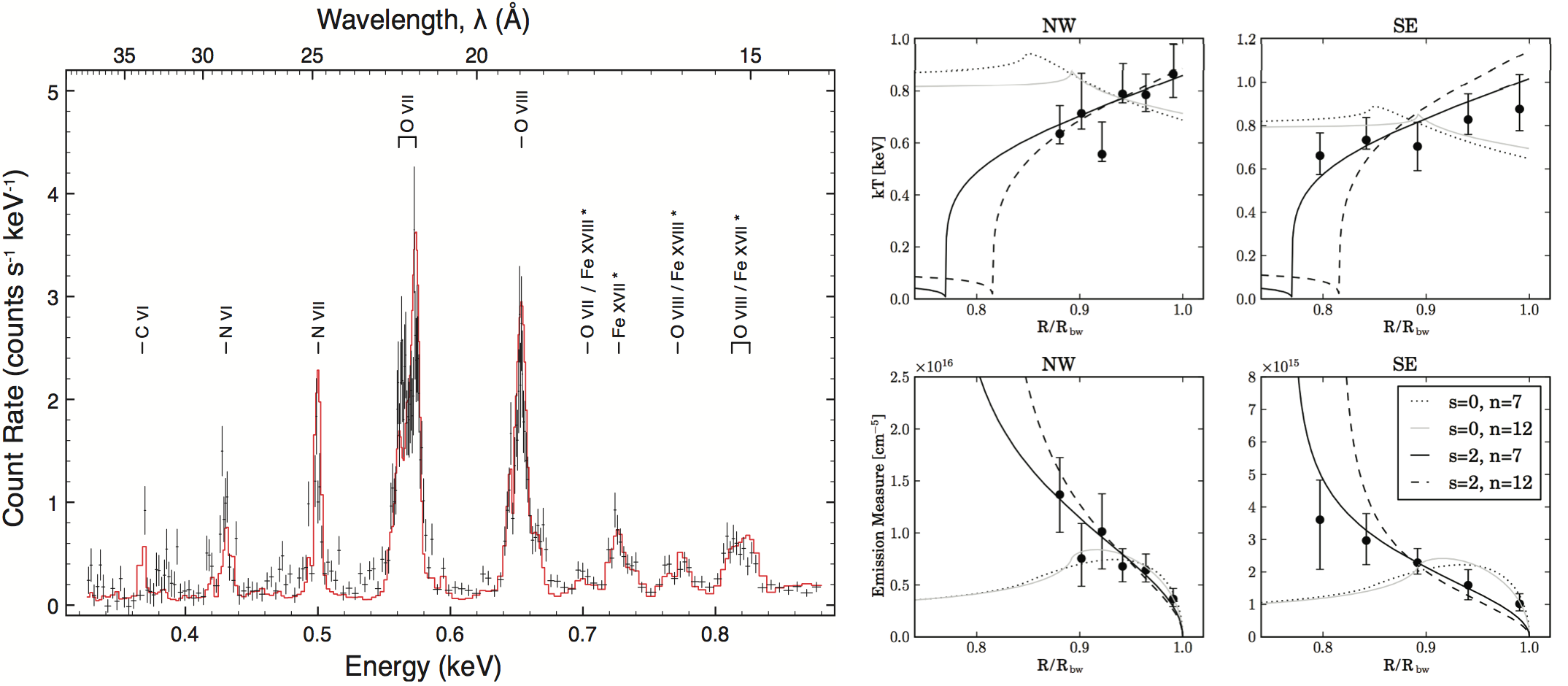}
 \caption{
Left: {\it XMM} RGS spectrum from G296.1$-$0.5 showing N and O line
features from shocked circumstellar wind. (From Castro et al.  2011.)
Right: Temperature and emission measure profiles of FS regions in
G292.0$+$1.8 compared with power law density models for uniform ($s
= 0$) and wind-like ($s = 2$) surroundings, and ejecta profiles
typical of Type Ia ($n = 7$ and CC ($n = 12$) events.
(From Lee et al. 2010.) [Figures reproduced by permission of the 
AAS.]
}
   \label{fig4}
\end{center}
\end{figure}

Detection of enhanced N in a shocked CSM environment is complicated
by both interstellar absorption and the modest spectral resolution
provided by typical X-ray CCD detectors. \xmm\ spectra of G296.1$-$0.5
from the MOS and pn detectors reveal emission from regions just
behind the FS that indicate a low column density and weak evidence
for an overabundance of N and an underabundance of O, as expected
from CNO-cycle products found in winds of massive stars. High
resolution spectroscopy using the {\it XMM} RGS (Figure 4, left)
confirm these results, clearly establishing the remnant as the
result of a CC event from a fairly massive progenitor
(Castro et al.  2011). Using the similarity solution of Chevalier
(2005), the inferred swept-up wind mass is $\sim d_2^{5/2} 19
M_\odot$, and the SNR age is $\sim 2800 E_{51}^{-1/2} d_2^{9/4}$~yr.

G292.0$+$1.8 is an O-rich SNR with an identified pulsar and PWN,
clearly establishing it as the result of a CC event.
X-ray spectral studies suggest a progenitor mass of $\sim 20 - 40
M_\odot$ (Hughes \& Singh 1994; Gonzalez \& Safi-Harb 2003; Park et
al. 2004), making it likely that the remnant has evolved in the
stellar wind density profile of its progenitor. \chandra\ studies
of the thermal X-ray emission from the outer regions of the SNR
shell (Figure 4, right) reveal a temperature and emission measure
structure consistent with predictions of the similarity solutions
from Chevalier (2005) for evolution in a medium with $\rho \propto
r^{-2}$, with a steep ejecta profile typical of CC events
(Lee et al. 2010). The overall kinematics are consistent with evolution
in an RSG wind comprising a mass of more than 15~$M_\odot$. More
recently, the same technique has been applied to the thermal
X-ray emission from Cas~A, revealing similar evidence of
a circumstellar environment dominated by an RSG wind (Lee
et al. -- these proceedings).

\section{Constraints on Particle Acceleration}
\noindent
Particle acceleration in SNRs has long been suggested as a primary
source for production of Galactic cosmic-rays, and X-ray measurements
have provided some of the most important constraints on this process.
While much of this evidence is provided by synchrotron radiation,
whose presence indicates the presence of multi-TeV electrons,
significant information is provided by the thermal X-ray emission
as well. In 1E~0102.2$-$7219, the temperature derived from X-ray
measurements is much lower than that implied by the shock velocity
determined from expansion measurements, even accounting for the
slowest possible equilibration between the electrons and ions (Hughes
et al. 2000b). This result is consistent with a picture in which a
significant fraction of the shock energy has gone into the acceleration
of particles instead of heating the gas.  In Tycho's SNR, the
separation between the FS and contact discontinuity (and also RS)
is much smaller than predicted from dynamical models of the SNR
evolution unless a significant amount of energy has been lost to
some nonthermal process, such as particle acceleration (Warren et
al.  2005). It is crucial to note that, since cosmic-ray protons
outnumber electrons by a factor of $\sim 100$, the significant
energy in relativistic particles inferred from these studies provides
strong evidence for acceleration of cosmic-ray {\it ions} in SNRs.

\begin{figure}[t]
\begin{center}
\includegraphics[width=1.00\textwidth]{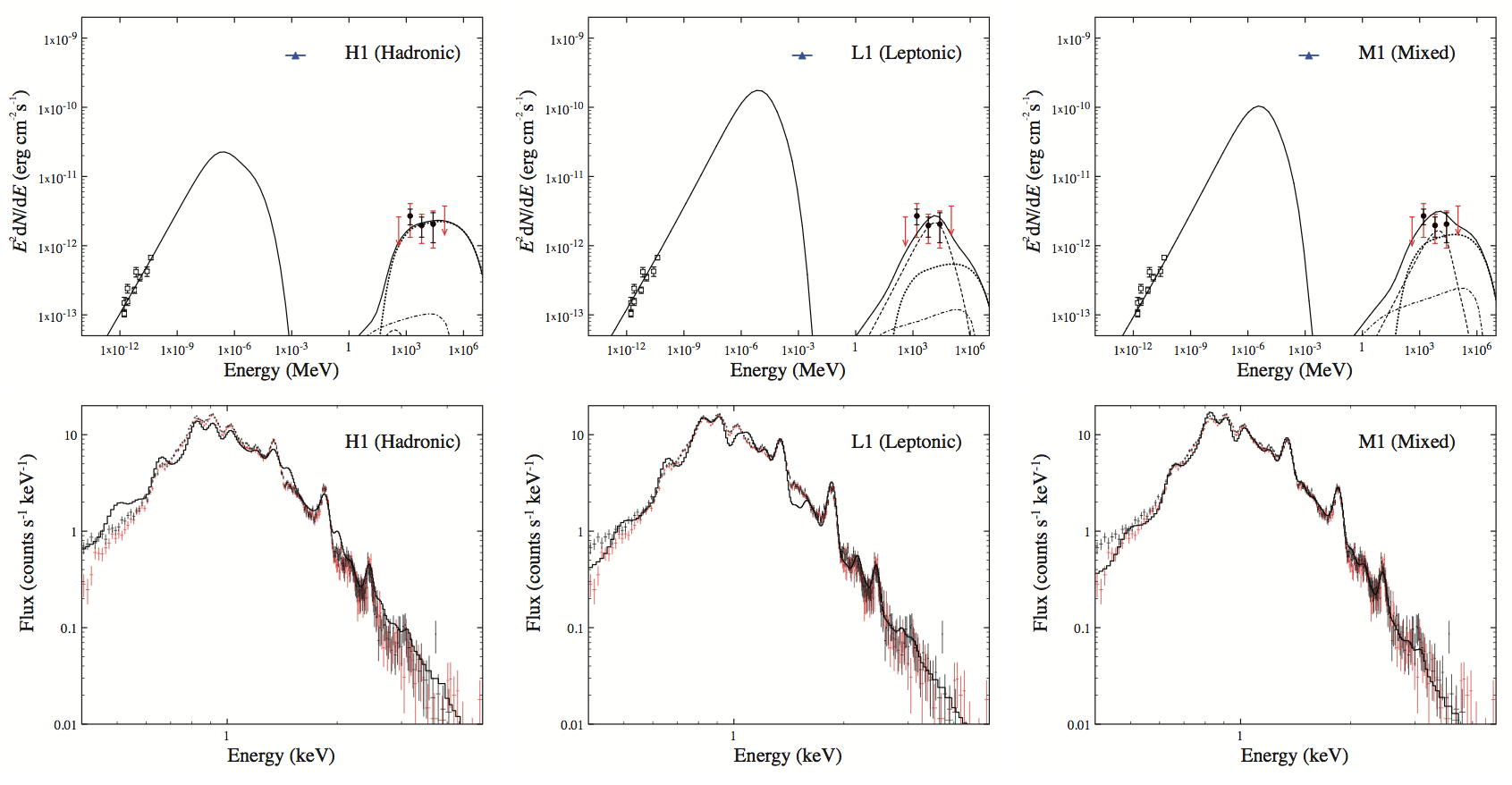}
 \caption{
Broadband (upper) and thermal X-ray (lower) spectra from CTB~109
along with models in which the $\gamma$-ray emission is dominated
by hadrons (left), electrons (center), and a mixture (right). The
thermal X-ray spectra provides a best fit for the mixed scenario.
(From Castro et al.  2012. Reproduced by permission of the AAS.)
}
   \label{fig5}
\end{center}
\end{figure}

Observations of $\gamma$-ray emission from SNRs provide additional
compelling evidence for particle acceleration. Because $\gamma$-rays
can be produced by both electrons (though inverse-Compton scattering
and nonthermal bremsstrahlung) and protons (though the production
of neutral pions, which decay to $\gamma$-rays), modeling of the
broadband emission is required to determine the origin of the
$\gamma$-rays, and thus the total efficiency with which the systems
are able to accelerate particles. In \RXJ, hydrodynamical modeling
of the SNR evolution that includes particle acceleration, and follows
the ionization history of the postshock gas, shows that the {\it
absence} of observed thermal X-ray emission places strong constraints
on the ambient density, effectively ruling out significant $\pi^0$-decay
$\gamma$-rays (Ellison et al. 2010, 2012) for scenarios with expansion
into a uniform medium or wind-driven cavity, although evolution in
a medium with dense clumps may provide sufficient target material
for a pions to provide a dominant contribution (Inoue et al. 2012).

In CTB~109, for which $\gamma$-ray emission is also observed,
self-consistent modeling (Castro et al. 2012) of the evolution can
produce broadband spectra that adequately reproduce the observed
radio and $\gamma$-ray emission for scenarios in which hadrons
dominate the $\gamma$-rays (Figure 5, left), or in which the
$\gamma$-rays originate primarily from electrons (Figure 5, center).
However, the thermal X-ray emission predicted by both of these
models fails to reproduce the observed \xmm\ spectrum. The high
density required for hadrons to dominate results in an overproduction
of high ion states for Mg and Si, while the low density required
for electrons to dominate results in an underprediction of the same
ion states. A mixed scenario in which the density is sufficiently
high for electrons and protons to contribute nearly equally to the
$\gamma$-ray flux yields excellent agreement with the thermal X-ray
emission.

\section{Conclusions}
\noindent
As indicated in this brief review, thermal X-ray spectra from SNRs
provide information on the nature, environments, and dynamical
evolution of these systems. Current X-ray observatories continue to
uncover new and important properties of both bright, nearby SNRs
and the fainter population within the Galaxy and beyond. Advances
in plasma codes, MHD simulations, and upcoming high spectral
resolution capabilities that will become available with {\it ASTRO-H},
hold particular interest for enhancing our abilities to probe the
detailed physics of SNRs and their evolution.

\bigskip \noindent 
This work was
carried out under support from NASA Contract NAS8-03060.

\end{document}